\documentclass[final,times,12pt]{elsarticle}

\usepackage{graphicx}   \usepackage{subfigure}
\usepackage{multirow}
\usepackage{multicol}
\usepackage{epstopdf}
\usepackage{amssymb}
\usepackage{tabularx}
\usepackage{caption}
\begin{document}
\raggedright 
\begin{frontmatter}

\title{An Analysis of Device-Free and Device-Based WiFi-Localization Systems}

\author{
\textit{Heba Aly, Dept. of Computer and Systems Engineering, Alexandria University, Egypt}\\[2mm]
\textit{Moustafa Youssef, Wireless Research Center, Egypt-Japan University of Science and Technology and Alexandria University, Egypt}
}

\begin{abstract}
WiFi-based localization became one of the main indoor localization
techniques due to the ubiquity of WiFi connectivity. However, indoor environments exhibit complex wireless propagation characteristics. Typically, these characteristics are captured by constructing a fingerprint map for the different locations in the area of interest. This fingerprint requires significant overhead in manual construction, and thus has been one of the major drawbacks of WiFi-based localization. 
In this paper, we present an automated tool for fingerprint constructions and leverage it
to study novel scenarios for device-based and device-free WiFi-based localization that are difficult to evaluate in a real environment. In a particular, we examine the effect of changing the access points (AP) mounting location, AP technology upgrade, crowd effect on calibration and operation, among others; on the accuracy of the localization system. We present the analysis for the two
classes of WiFi-based localization: device-based and device-free.
Our analysis highlights factors affecting the localization system accuracy, how to tune it for better localization, and provides insights for both researchers and practitioners.
\end{abstract}


\end{frontmatter}

\section*{Introduction}
Due to recent advances in wireless networking, WiFi-based localization has been attracting significant attention. WiFi deployments are ubiquitous in public and private places including offices, malls, and hospitals. WiFi-based localization techniques use the existing ubiquitous WLANs to provide accurate indoor localization without any additional hardware. It can be classified into two categories: device-based~\cite{radar,horus}, and device-free techniques~\cite{Challenges,nuzzer,rasid,kalman}.
Device-based systems track a WiFi-enabled device such as a cell-phone, based on the received signal strength (RSS) at this device; while device-free systems track entities that do not carry any devices based on their effect on the RSS at the infrastructure devices. A typical device-free system will consist of one or more signal receivers which are called monitoring points (MPs) such as laptops; signal transmitters such as access points (APs); and also an application server, which is usually one of the monitoring points, to collect data from monitoring points. Applications for the device-free systems include intrusion detection, smart homes, and sensor-less sensing.

\begin{figure}[!t]
  \centering
     \subfigure[Active]{
      \includegraphics[width=0.46\columnwidth]{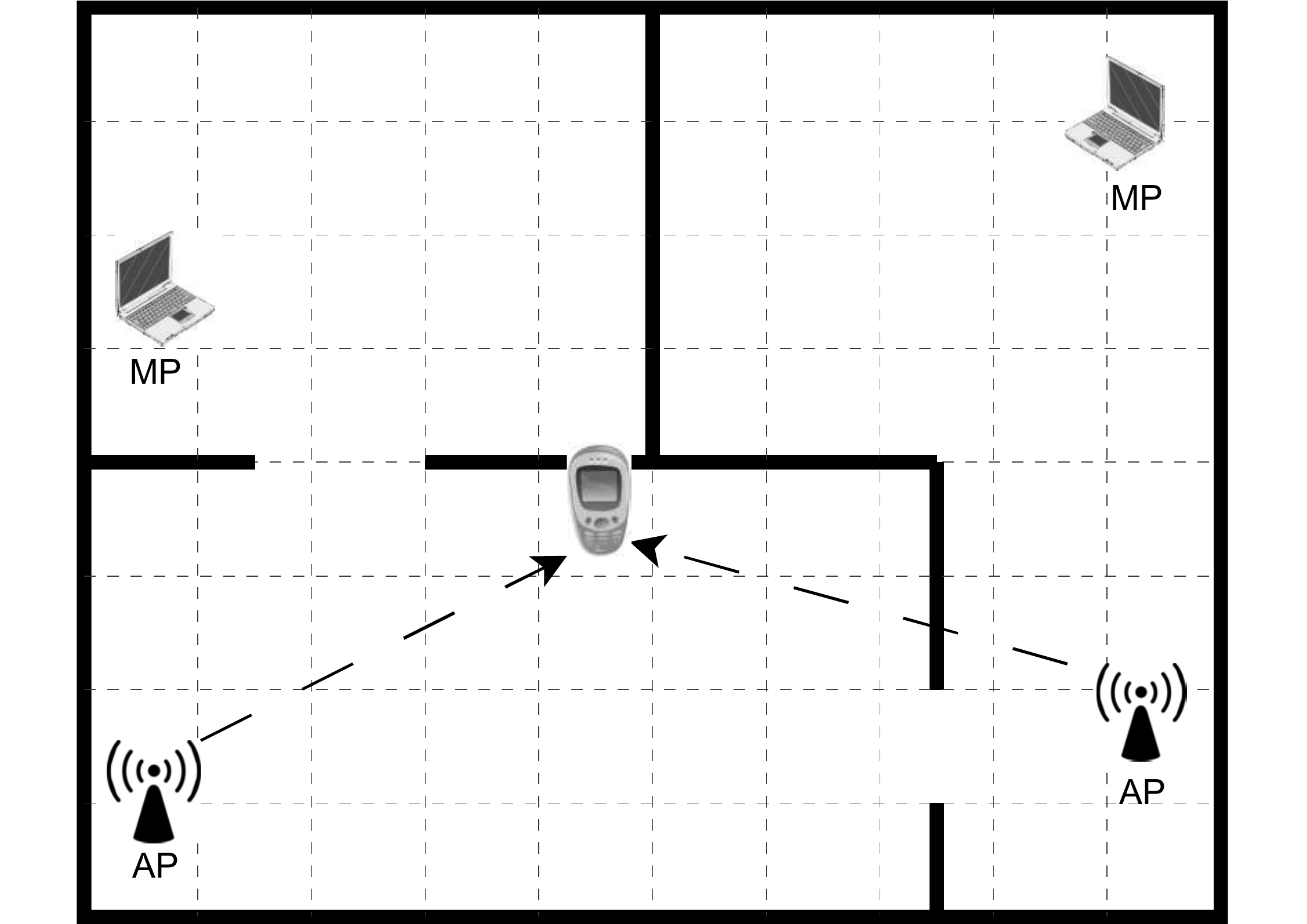}
      \label{fig:rm_active}
    }
    \subfigure[Passive]{
      \includegraphics[width=0.46\columnwidth]{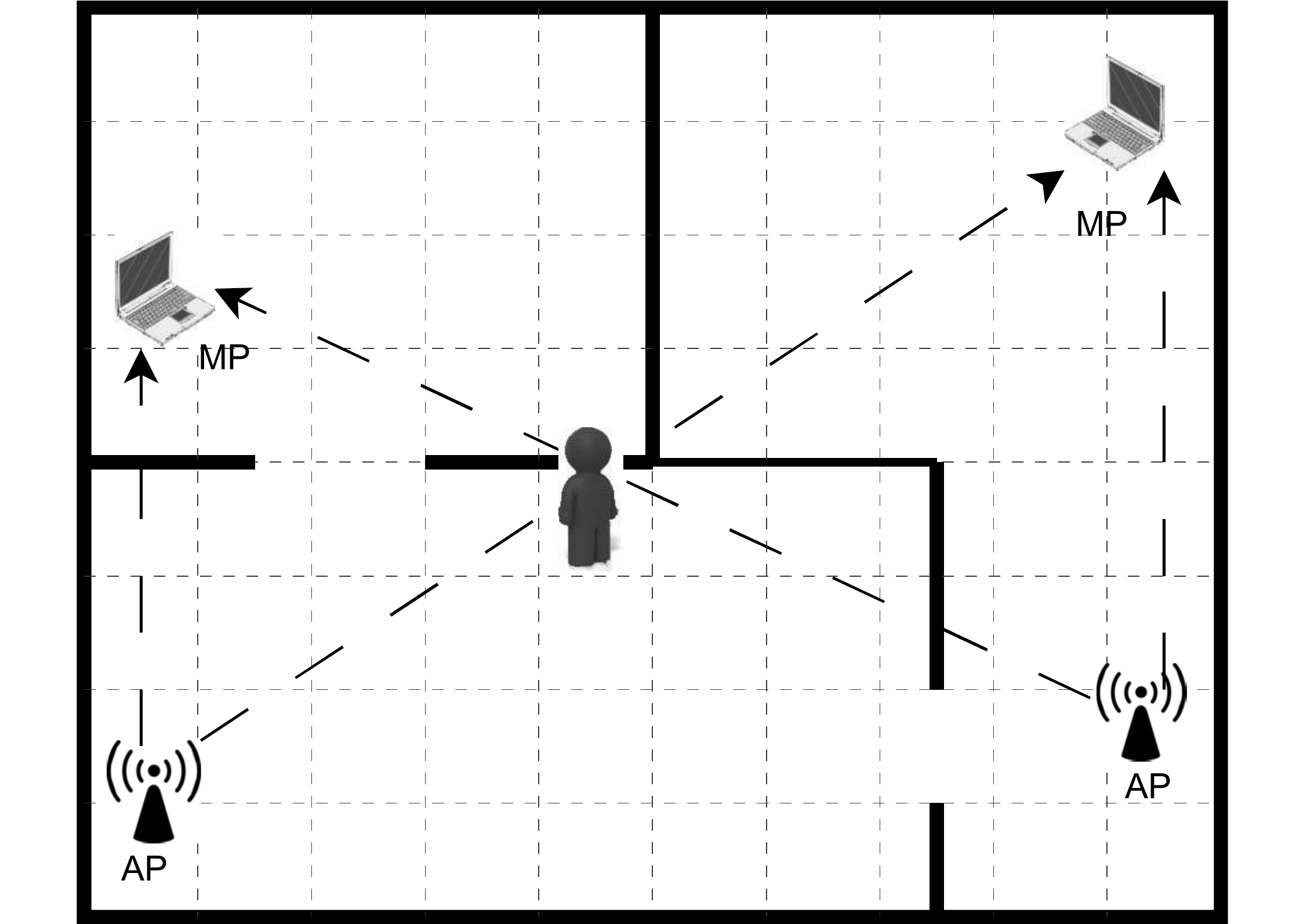}
  \label{fig:rm_passive}
    }
  \caption{Difference between device-based and device-free radio-map
construction.}
  \label{fig:db_df_rm}
\end{figure}

To overcome the complex propagation characteristics of WiFi signals in indoor environments~\cite{sscompwlan}, typically both device-based and device-free systems require a calibration phase to construct a fingerprint or a radio-map that stores the RSS characteristics at different locations in the area of interest. Device-based systems use active-radio maps, where each stream represents the signal strength from an AP to the tracked device; while device-free systems use passive-radio maps, where each stream represents the signal strength from an AP to a MP capturing the effect of the tracked entity on the fixed streams. Figure~\ref{fig:db_df_rm} explains the difference between active and passive radio maps construction.

Traditional methods of radio-map construction require the use of manual calibration, which is a tedious and a time consuming process. Therefore radio map construction has been based on simple scenarios, usually involving one entity in a specific environment.

Received signal strength (RSS) is sensitive to different system configurations, environmental parameters and human presence\cite{radar,kaemarungsi2004properties,kaemarungsi2012analysis,el2011impact}. Analyzing the RSS characteristics can help ease the localization system design and improve the localization accuracy. Hence, different studies were conducted to study the factors affecting the device-based and the device-free localization systems.  For example, in~\cite{kaemarungsi2004properties}, the authors investigated the effect of user's presence and user orientation on the RSS of a device-based system. They also studied the effect of having multiple APs and the relation between their RSSs. In~\cite{kaemarungsi2012analysis}, they further investigated other parameters affecting the RSS such as the effect of the hardware used (e.g. WLAN cards from different vendors) and the technology used (IEEE 802.11/b and 802.11/g) for device-based localization. In~\cite{el2011impact}, authors showed the effect of human movement on the variance of the RSS and in~\cite{kosba2009analysis} authors investigated the effect of temporal and spatial changes in the environment on the accuracy of the device-free localization system and also the effect of the APs and MPs locations.

Different from those experiments, we use AROMA~\cite{aroma,aromaglobecom}: a state-of-the-art accurate system for automatic generation of WiFi-based radio-maps, to analyze different what-if scenarios for both device-based and device-free WiFi-based localization that was hard to evaluate in real environments. In particular, we study the effect of upgrading the hardware, APs mounting locations, significant differences between the calibration and operation environments, and outsider's effect on device-free localization. Through these scenarios we show different factors affecting device-based and device-free localization, giving new insights into WiFi-based localization that can be leveraged by users, developers, and researchers.

The rest of the paper is organized as follows: In the next section, we start by discussing the related work and give an overview of the AROMA system. Then, we present our device-based and device-free scenarios showing their effect on the RSS and the system accuracy in the two following sections. Finally, we conclude our work highlighting our recommendations for better localization.
\section*{Related Work}\label{sec:relwork}
The proliferation of WLANs induced a great interest in leveraging them to provide ubiquitous localization. The WLAN localization techniques are mainly classified into two types: device-based and device-free. Typically, the device-based localization techniques localize cell-phones by matching the observed RSS readings (from the phones) against a pre-calibrated finger-print map. The finger-print map can be constructed either using deterministic approaches, e.g. \cite{radar}; or probabilistic approaches, e.g. \cite{horus}. The RADAR\cite{radar} system was one of the first systems that used a deterministic fingerprint for each location. Since then several schemes have improved upon it, e.g. the Horus\cite{horus} system, which employed a stochastic description of the RSS fingerprint map and a maximum likelihood based approach for localization. The device-based approach has been more intensively studied in comparison with the device-free class, with even some commercial applications. Recently, new approaches that reduce the calibration overhead were introduced. For example, in \cite{wang2012no} the system depends on crowdsourcing the WLAN fingerprint construction.

The device-free concept have been an active research area since its introduction \cite{Challenges}. The technique was further investigated in \cite{moussa2009smart,kosba2009analysis,el2010propagation,seifeldin2010deterministic,kosba2012robust} to provide accurate indoor localization for a single entity in a typical home environment.
In \cite{sabek2012multi,sabek2012spot}, authors provided accurate localization for more than one entity. Recently, more sensing dimensions were explored. In \cite{kassem2012rf,al2012rf}, authors applied the DfP concept for vehicles and transportation mode detection. In\cite{abdel2013monophy}, authors investigated the effect of using physical layer information to minimize the number of monitoring points and access points used. They could provide accurate localization information using only one stream.
The DfP concept was extended to detect different activities performed in an area of interest\cite{scholz2011challenges,hong2013ambient,sigg2013rf}, where authors explored the detection of different activities performed by a single entity.

Analyzing the localization system characteristics is mandatory for a better and easier system design. For example, if there is a set of guidelines, this can ease the design and configuration of the localization system. Therefore researchers developed interest in analyzing the RSS characteristics and the different factors affecting the localization systems. Throughout this section, we discuss the different conducted studies to analyze both types of the WiFi localization systems. Then, we give an overview of the AROMA system that we used to provide our insights.

\subsection*{Analysis for Device-based Systems}

Different studies were conducted to analyze the device-based systems, e.g. \cite{radar,kaemarungsi2004properties,kaemarungsi2012analysis}. In~\cite{radar}, authors performed basic analysis on the signal strength at different locations for different user orientations. They pointed out that the user orientation has a significant impact on the signal strength. They also studied the impact of the number of fingerprint data points (radio map density) on the system accuracy and the number of samples collected per location. They recommended choosing uniformly distributed locations across the area of interest and concluded that only small number of samples per location is enough. They further studied the effect of the number of nearest neighbors in their classification algorithm and the radio propagation models.

In~\cite{kaemarungsi2004properties}, authors investigated the properties of the received signal strength reported by IEEE 802.11b wireless network interface cards and analyzed the data to understand the underlying features of location fingerprints. They also analyzed the effect of the user's presence on a single RSS set, its statistical properties (e.g. its distribution and the time-of-day dependency), and the properties of RSS values from multiple APs. Also, in~\cite{kaemarungsi2012analysis}, they further investigated other parameters affecting the RSS such as the effect of the hardware used (e.g. WLAN cards from different vendors) and the technology used (IEEE 802.11/b and 802.11/g) while using the same frequency band (2.4 GHz).

Choice of APs locations does not only affect the WLAN coverage but it affects the system localization accuracy too. In\cite{chanproperties} authors investigated the influence of channel interference, the channel assignment of APs, the distribution of received signal strength (RSS) values, the variation of coverage, and
distances between APs on the performance of the localization system.

Different from these studies, we analyze the effect of the APs height on the system, the effect of the technology used for different frequencies (2.4 GHz and 5 GHz), the effect of crowd presence and we also investigated the effect of changes in the environment between the training and testing and how to mitigate it. These scenarios are harder to evaluate in a real environment due to the labour and monetary costs.
\begin{figure}[!t]
	\centering
	\includegraphics[height=8cm]{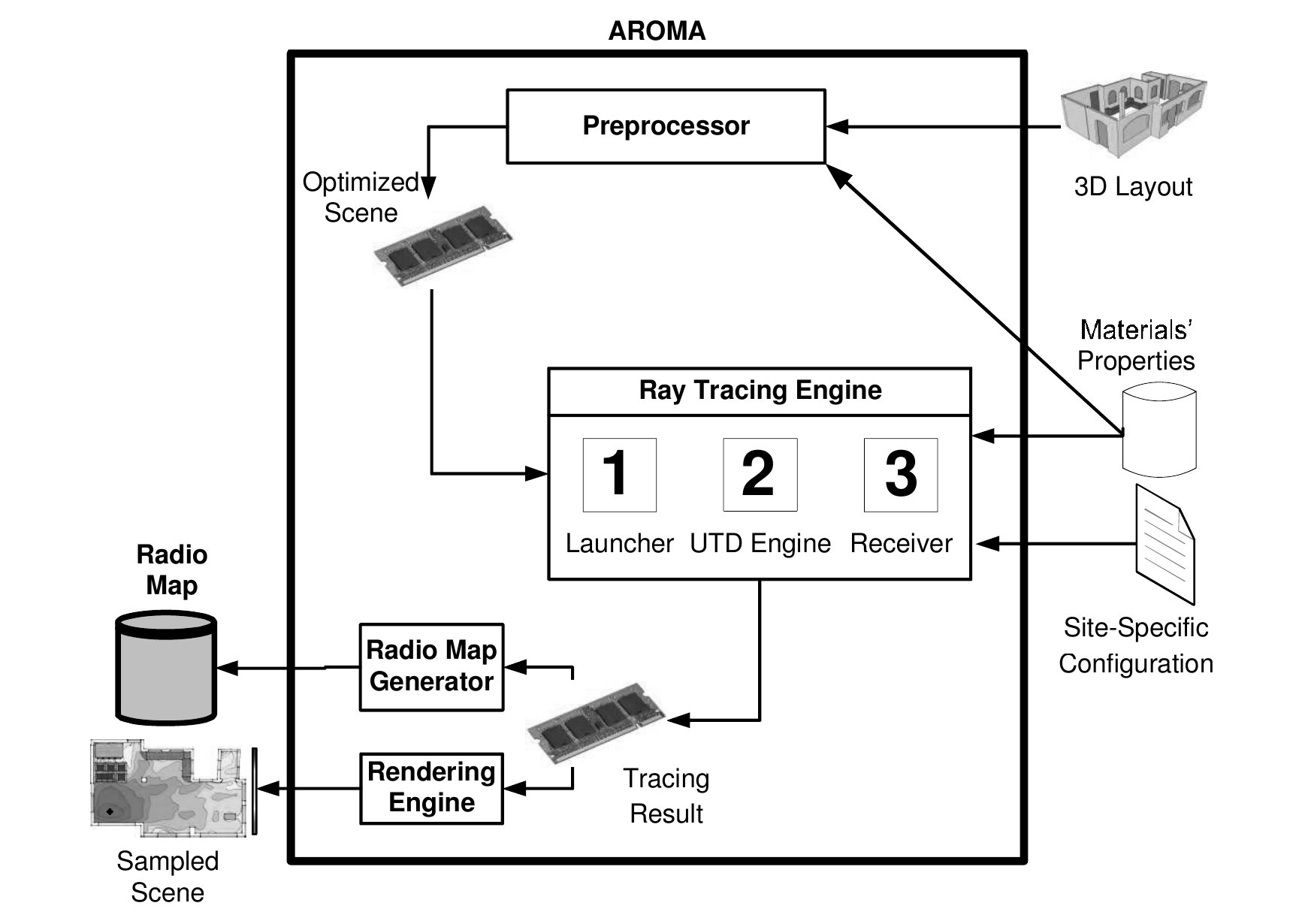}
	\caption{The AROMA system architecture.}	\label{fig:aroma_arch}
\end{figure}
\subsection*{Analysis for Device-Free Systems}
Device-free localization depends on the human effect on a fixed set of MPs and APs streams. Analyzing the effect of the human presence and human motion on the RSS streams is crucial for any device-free localization system.
In\cite{moussa2009smart}, authors showed the effect of changes in the real environment on the device-free localization system; the RF signals were affected by changes such as the frequency ranges of the data network. They showed the spatial and temporal variability of the signals due to human motion and multi-path effect respectively.
In~\cite{kosba2009analysis}, authors investigated the effect of temporal and spatial changes in the environment, and number of APs and MPs steams on the accuracy of the localization. They also studied the effect the APs and MPs locations. Also in~\cite{el2011impact}, authors investigated the effect of human movement on the variance of the RSS.

In our study, we investigate the effect of the APs mounting location (for the same AP placement, we study the effect of its mounting height), the effect of changing the technology used, and the effect of outsiders on the system.

\subsection*{The AROMA System}\label{subsec:aroma}

AROMA~\cite{aroma,aromaglobecom} is the state of the art system for automatic radio map generation for both device-based and device-free localization systems. It is the first system to consider radio map generation for device-free systems and the first to consider human effects in the device-based systems. AROMA utilizes site-specific ray tracing, augmented with the uniform theory of diffraction (UTD)~\cite{UGTD}, to predict the RF propagation in a 3D site. It models the human body as a metallic cylinder~\cite{ghaddar2004modeling}. Figure~\ref{fig:aroma_arch} shows the architecture of the AROMA system.

AROMA takes as an input a 3D model of the site of interest
and the site-specific configurations (the APs and the receivers locations and antenna configurations). It comes with a built-in database of the values of the RF propagation
properties of common building materials such as bricks and
concrete. The user has the options of using this database or
providing customized values using the user interface tool.
The Ray Tracing Engine is the core of the AROMA system
and is composed of three modules: the Ray Launcher, the
UTD Engine, and the Ray Receiver. The Ray Launcher is
responsible for emitting electromagnetic waves from APs~\cite{seidel1994site}.
Ray tubes propagate in the complex indoor environment
experiencing reflection, transmission, or diffraction causing
multipath fading. Rays are traced up to a user-defined depth;
each time the ray makes an interaction with the environment,
its depth is incremented. The tracing of a ray ends if the
depth reaches the maximum user-defined value or the power
associated with the ray decreases below a defined minimum
value. The UTD Engine handles the changes in the electric
field associated with the ray tubes resulting from their interactions
with the environment. These changes are modeled with
Geometric Optics (GO) augmented with the Uniform Theory
of Diffraction (UTD). The Ray Receiver finds the contribution
of each ray to the final RSS at the receivers, using the reception
sphere model~\cite{seidel1994site}. Different optimization techniques are used
for efficient computations.

The AROMA system was validated in actual testbeds and showed that the predicted signal strength differs from the measurements by a maximum average absolute error of 2.77 dB  and a maximum localization error
of 3.13m for both the device-based and device-free cases.
\newline

In \cite{aly2013new}, we leveraged the AROMA system to study different scenarios and provide our recommendations for better device-free localization systems. Through this paper, we extend our study to include device-based localization systems and give our recommendations for better device-free and device-based localization.
\section*{Device-based Scenarios}\label{sec:db_scenarios}
In this section, we analyze factors affecting the device-based localization systems through different scenarios. In particular, we study the effect of the APs mounting locations, upgrading the APs hardware (changing the operation frequency), and significant changes between the calibration and operation phase. We end the section with a summary of our findings.
We start by describing the experimental testbed and evaluation metrics.
\begin{figure}[!t]
	\centering
	\includegraphics[width=0.7\linewidth]{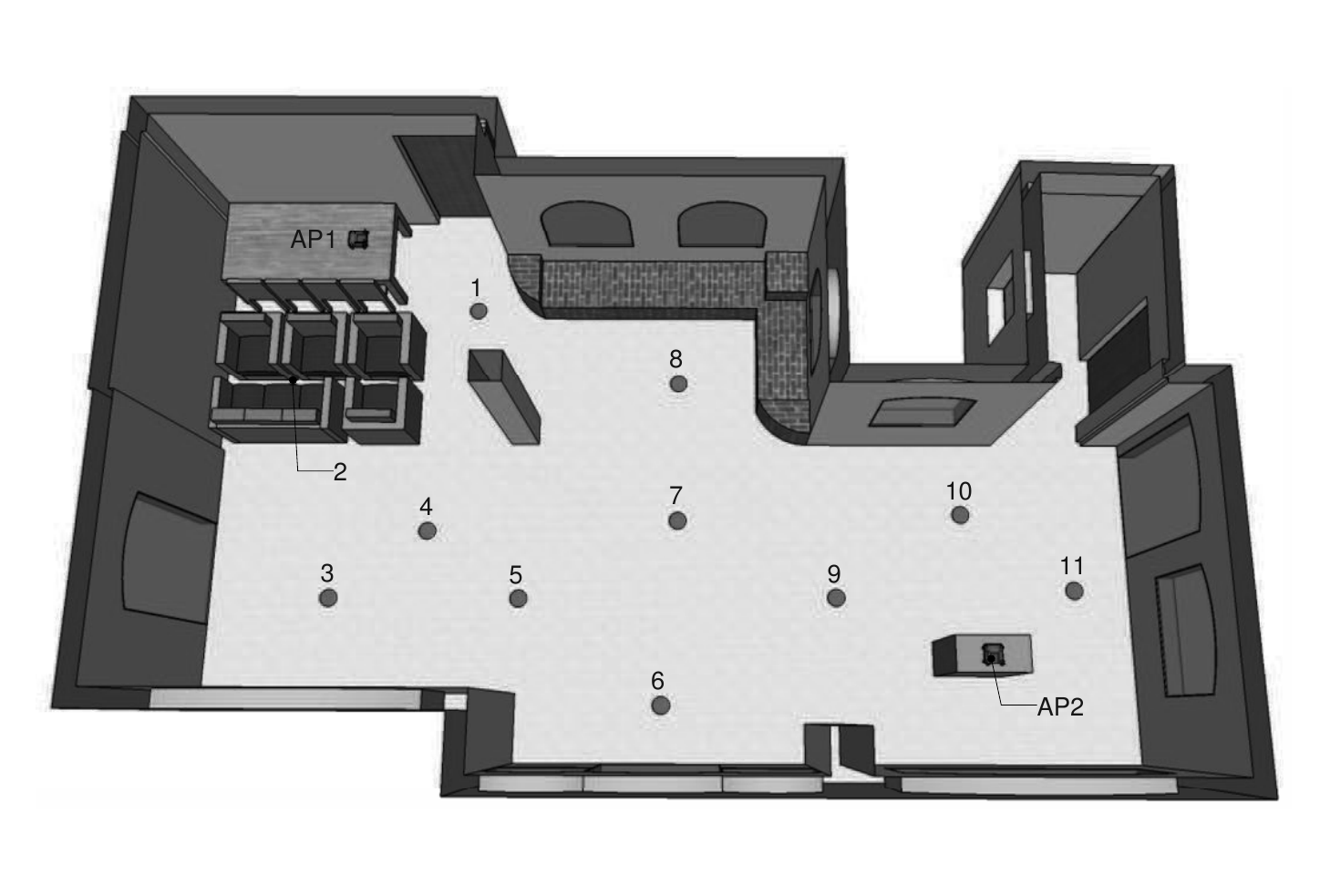}
	\caption{Device-based experiments layout. The figure highlights locations of APs and radio map locations.}
	\label{fig:DeviceBased_Labeled}
\end{figure}
\begin{table}[!t]
		\centering
\caption{Default APs configuration}
		\begin{tabular}{|c|c|}
		\hline Trans. power ($P_{t})$ & $2$ mW \\
		\hline Antenna gain ($G_{max}$) & $3$ dBi \\
		\hline Frequency ($f$) & $2.4$ GHz. \\
		\hline Antenna type & Isotropic \\
		\hline Location & Wall-mounted ($1.5~m.$) \\
		\hline Celling height & $2.7~m.$ \\
		\hline
	\end{tabular}
	\label{table:db_ap_config}
\end{table}
\subsection*{Experimental Testbed and metrics}
We used a 3D model of a typical apartment with an area
of $700 ft^2$($66~m^{2}$). The environment is composed of different
materials and contains some furniture. The radio-map locations
are marked in Figure~\ref{fig:DeviceBased_Labeled}; they are eleven different locations covering the entire area of the apartment. The default APs
configurations are summarized in Table~\ref{table:db_ap_config}.
To study the effect of the different scenarios, we used two
metrics: the change in the RSS and the change in the average
localization error in meters when using a nearest neighborhood
classifier, e.g.~\cite{radar}.

\begin{table}[!t]
	\centering
\caption{Average localization error in meters for different device-based scenarios. All experiments were trained with just one person (no crowd).}
\resizebox{1\linewidth}{!}{
	\begin{tabular}{|l|l||l|l|}
		\hline {\bfseries Experiment} & {\bfseries Acc.} (m) & {\bfseries Experiment} & {\bfseries Acc.} (m)\\\hline
		\hline {\bfseries Base exp.}  (params. in Table~\ref{table:db_ap_config}) & 1.84 & Exp. 1: Ceiling-mounted APs & 1.00\\
		\hline Exp. 2: Freq. 5.7 GHz & 1.55 & Exp. 3: Crowd around AP1 & 2.36\\
		\hline Exp. 3: Crowd around AP2 & 3.06 & Exp. 3: Crowd around both APs & 3.03\\
		\hline Exp. 3: Party, wall-mounted APs & 2.02 & Exp. 3: Party, ceiling-mounted APs & 1.64\\
		\hline
\end{tabular}
}
	\label{table:dbloc_acc}
\end{table}
\begin{figure}[!t]
	\centering
	\includegraphics[width=0.6\linewidth]{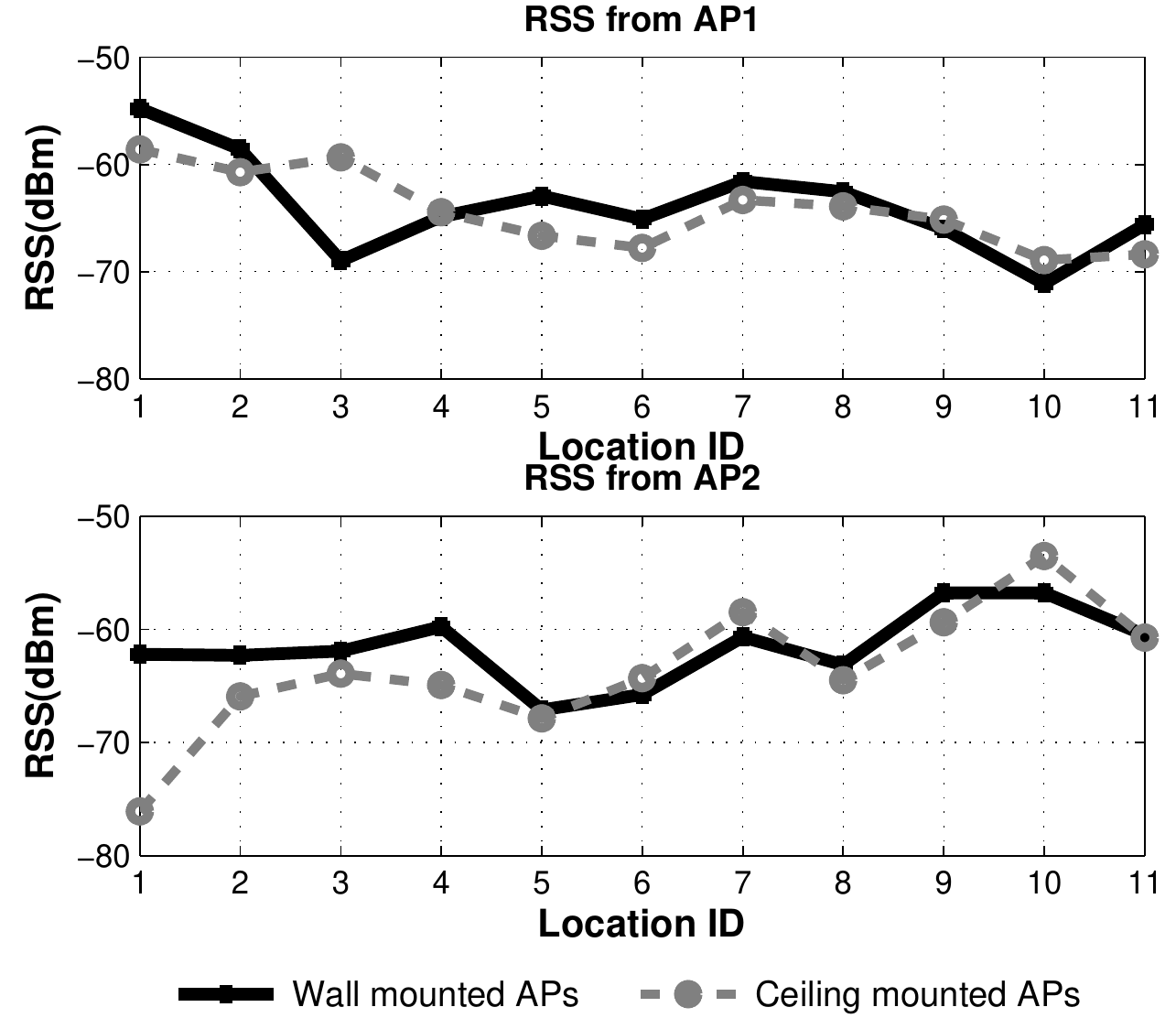}
	\caption{Effect of APs mounting location on the RSS from two APs at different
radio-map locations (IDs are in Figure~\ref{fig:DeviceBased_Labeled}).}
	\label{fig:db_apheight}
\end{figure}
\subsection*{Exp. 1: AP Mounting Location}
APs are usually mounted either on walls or on the ceiling.
This experiment investigates the effect of placing the APs at the ceilings or on walls in
the same environment. Figure~\ref{fig:db_apheight}
shows the effect on the RSS while Table~\ref{table:dbloc_acc} shows the localization accuracy.\\

The results show that in general, ceiling-mounted APs lead
to a weaker signal. This is expected due to the longer distance
traveled by the signal. Note that wall-mounted APs make
the signal interact with more objects, and hence reduce the RSS due to attenuation. However, the results show that the
distance effect is the dominating effect in most cases. This
larger number of interactions also explains why the variance
of the RSS is less in the case of the wall-mounted APs. This
lower differentiation between locations, in terms of RSS, leads
to lower localization accuracy as shown in Table~\ref{table:dbloc_acc}.
\newline
The locations that had the largest change in RSS are
those close to objects (e.g. Location 1); due to the complex
wireless propagation and interaction with nearby objects, those
locations were the most affected.

\subsection*{Exp. 2: Updating the AP H/W (Changing the Operating
Frequency)}
\begin{figure*}
\begin{minipage}[H]{0.55\linewidth}
	\centering
	\includegraphics[width=\linewidth]{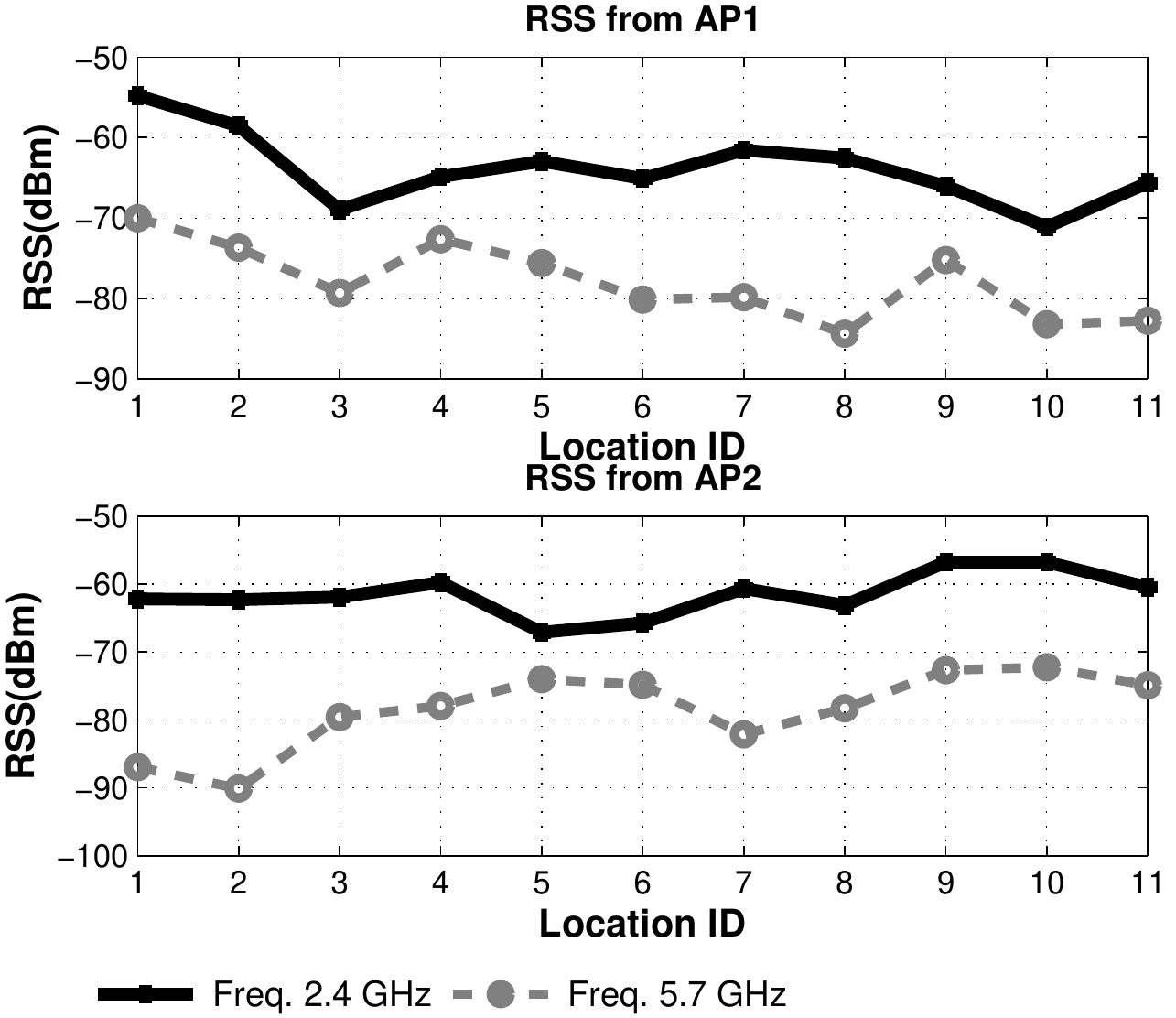}
	\caption{Effect of switching the frequency from 2.4 GHz to 5.7 GHz.}
	\label{fig:db_freq}
	\end{minipage}
\hspace{0.01\linewidth}
\begin{minipage}[H]{0.42\linewidth}
\centering
	\subfigure[Crowd around both APs.]{
		\includegraphics[width=0.8\linewidth,height=2cm]{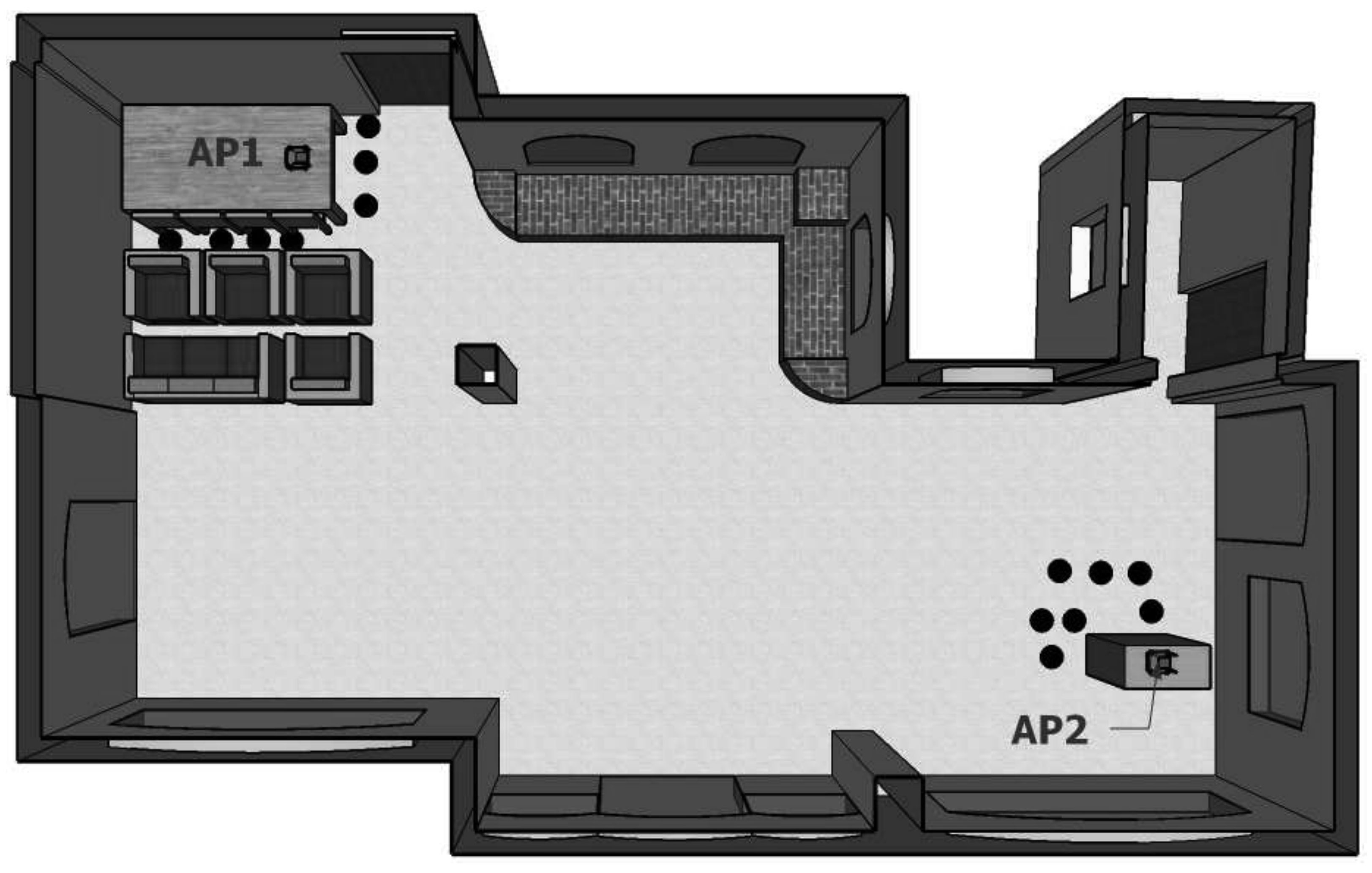}
		\label{fig:crwd_both}
	}
	\subfigure[Crowd in the party scenario.]{
		\includegraphics[width=0.8\linewidth,height=2cm]{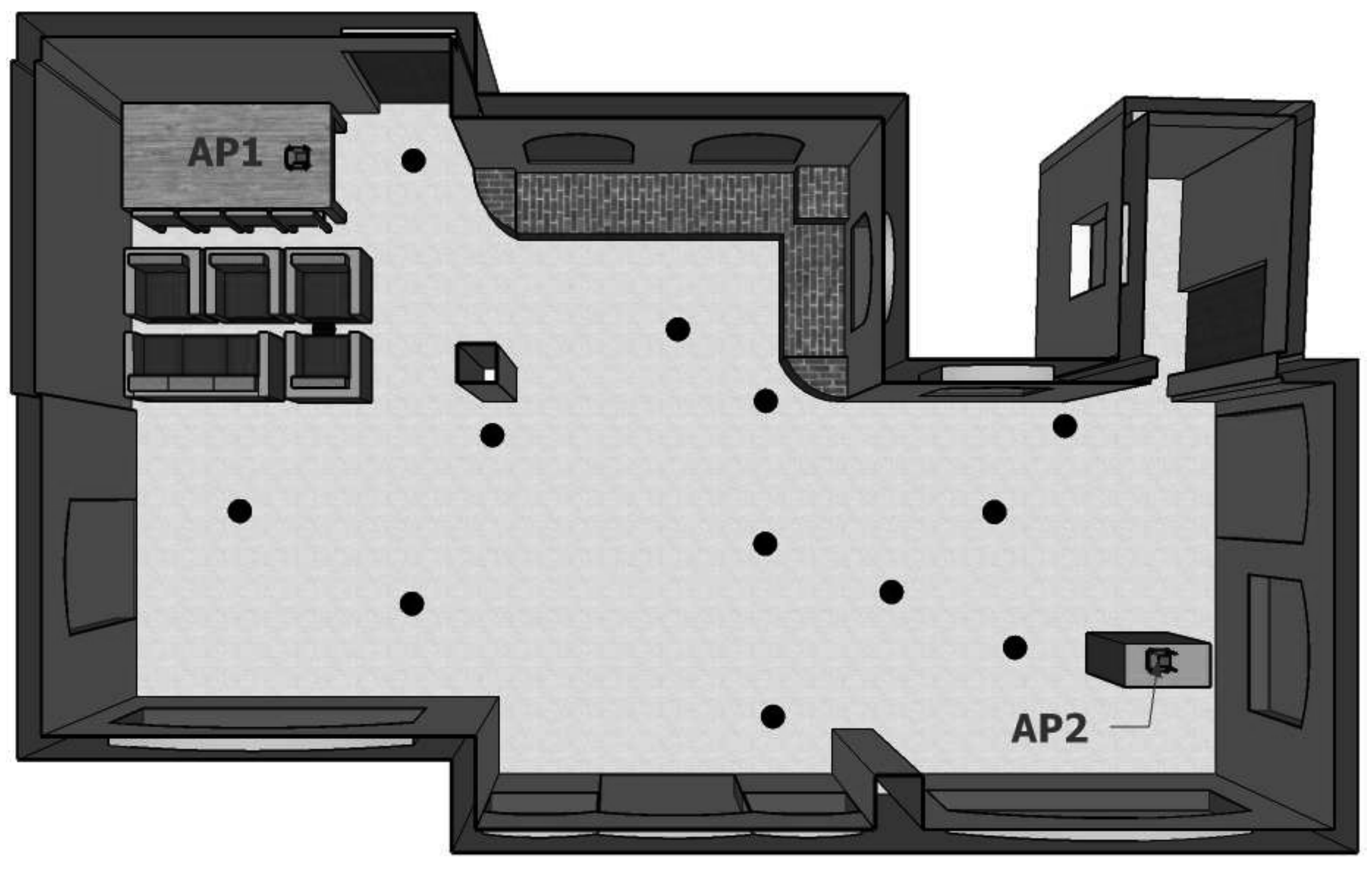}
		\label{fig:dcrwd}
	}
	\caption{The crowd patterns used: a crowd distributed around the apartment,
and the crowd around APs. Entities locations are highlighted in black.}
		\label{fig:dbcrowd}
\end{minipage}

\end{figure*}
Different WiFi technologies operate on different frequency
bands; e.g. IEEE 802.11/b/g use the 2.4 GHz band while
802.11/a uses the 5 GHz band. This experiment studies the
effect of changing the operating frequency. Figure~\ref{fig:db_freq} shows
the effect on the RSS while Table~\ref{table:dbloc_acc} shows the localization
accuracy (2.4~GHz is the band in the base experiment).
The figure shows that, due to the shorter wave length, the
higher frequency lead to a higher attenuation and a weaker
RSS. This leads to a lower coverage area, higher variance
within the same area, and hence better localization as shown
in Table~\ref{table:dbloc_acc}.
\begin{figure}[!t]
	\centering
	\includegraphics[width=0.6\linewidth]{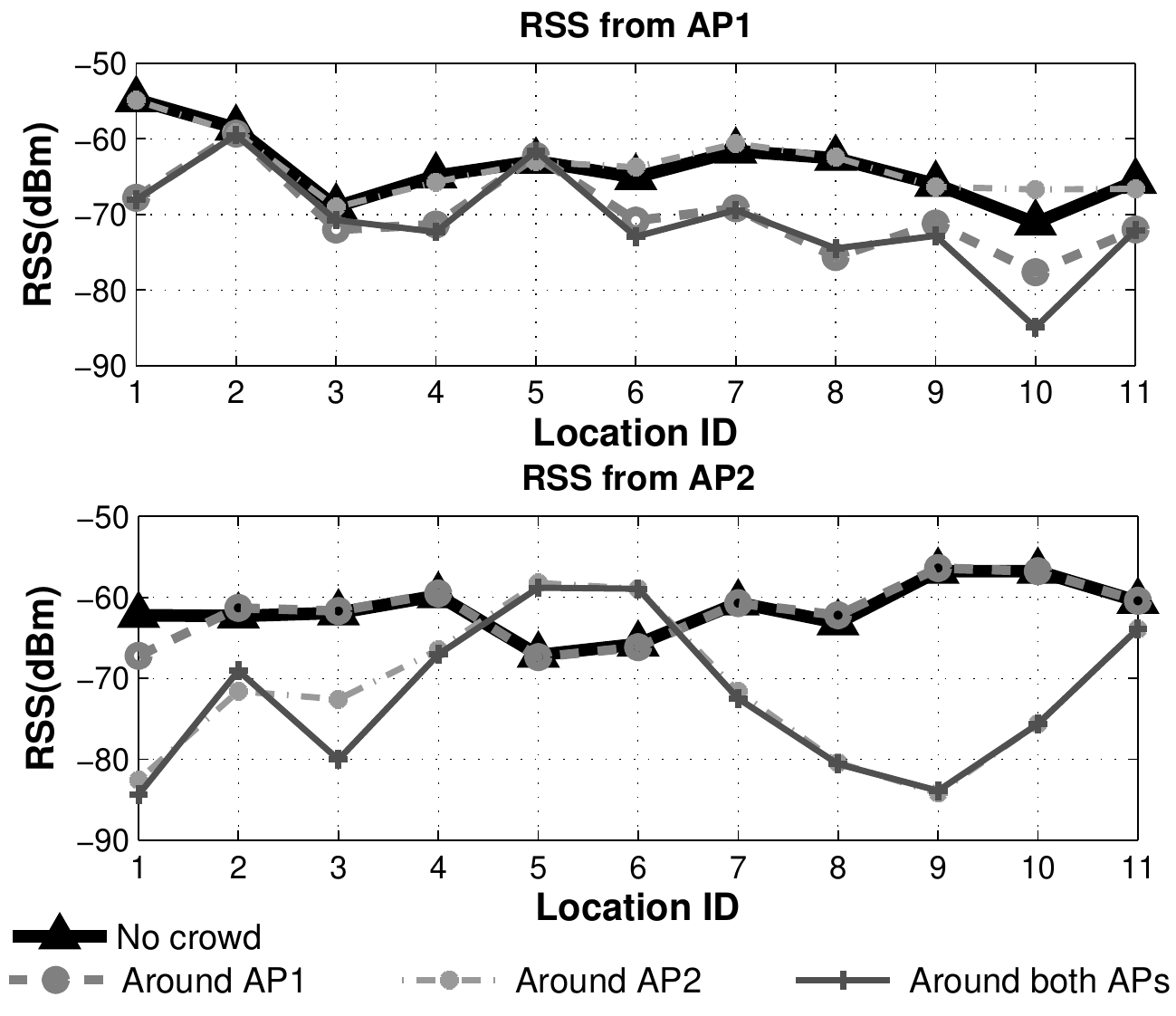}
	\caption{Effect of having a crowd around AP1, a crowd around AP2 and a crowd around both APs (all for wall-mounted APs).}
	\label{fig:both2}
\end{figure}
\subsection*{Exp. 3: Significant Changes between Training and Operation (Crowd Scenario)}
Device-based localization systems are usually used in environments with different crowd patterns and it is hard to make any assumption on how this crowd will be scattered in the area of interest. A human body cutting any path between the transmitter and receiver causes an attenuation to the signal~\cite{kaemarungsi2004properties}. Through this experiment we investigate the effect of a crowd on the RSS and the localization accuracy through four different crowd patterns: (1) a crowd around AP1, (2) a crowd around AP2, (3) a crowd around both APs (Figure~\ref{fig:crwd_both}), and (4) a crowd distributed in the whole area simulating a party scenario as in Figure~\ref{fig:dcrwd}. These patterns represent the extremes for the crowd attenuation and a general case (the party scenario). We analyze these four scenarios using both ceiling-mounted and wall-mounted APs. Since it does not make sense to surround a ceiling-mounted AP with a crowd, we only analyze the ceiling-mounted APs with a party scenario (Figure~\ref{fig:dcrwd}).\\
\begin{table}[!t]
	\center
		\caption{Average localization error in meters for different {\bfseries party} scenarios.}
	\resizebox{0.95\linewidth}{!}{
	\begin{tabularx}{\linewidth}{|X||X|X|}
		\hline \multirow{2}{*}{{\bfseries Experiment}} & \multicolumn{2}{|c|}{{\bfseries Accuracy (m)}} \\\cline{2-3} & Trained with crowd & Trained with no crowd      \\
		\hline Wall Aps & 2.35 & 2.02 \\
		\hline Ceiling APs & 2.09 & 1.64 \\
		\hline
	\end{tabularx}
	}
		\label{table:dbloc_acc_1}
\end{table}

\begin{figure*}
\begin{minipage}[H]{0.5\linewidth}
\includegraphics[width=\linewidth]{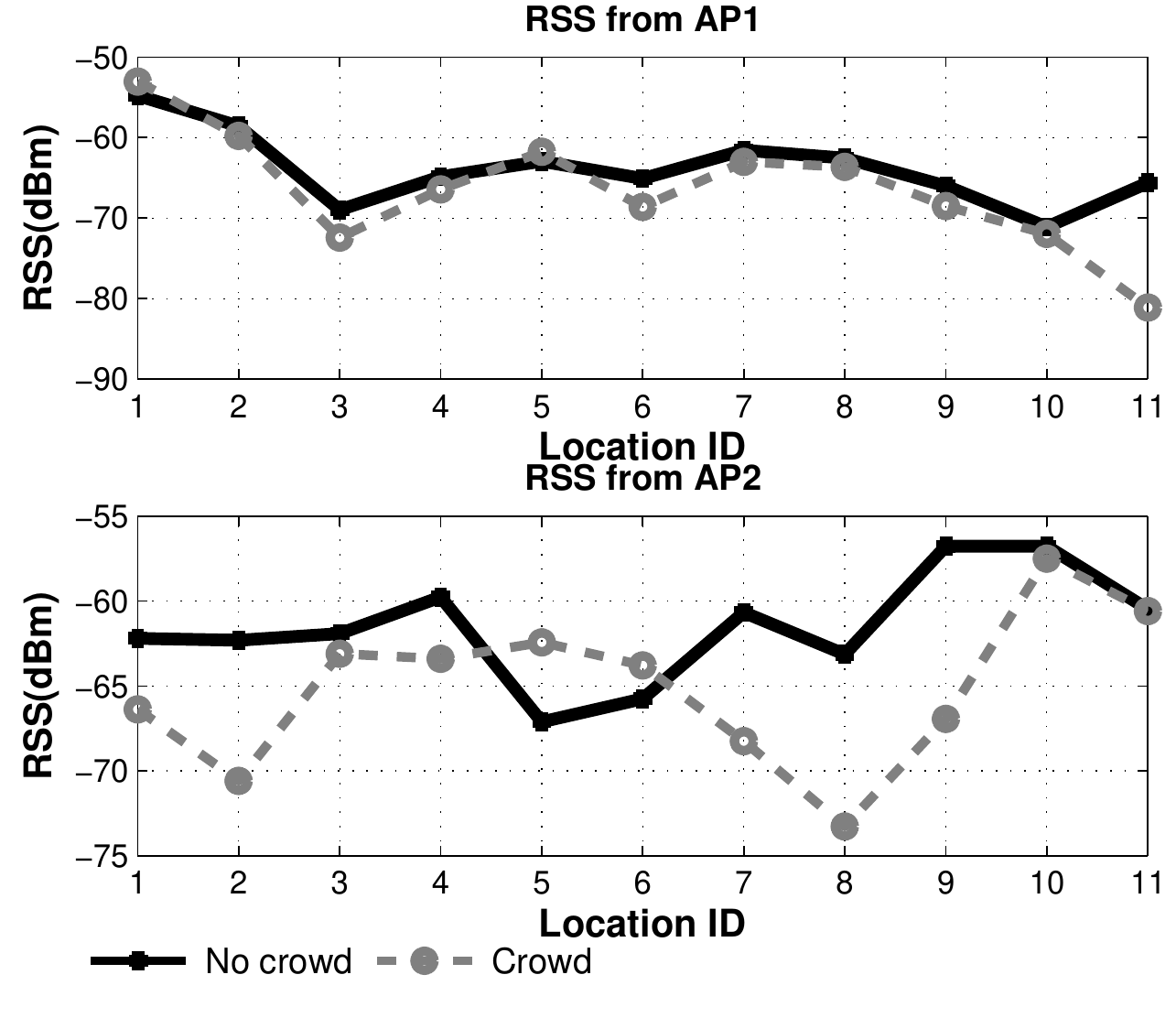}
	\caption{Effect of the {\bfseries party} crowd when using wall-mounted APs.}
	\label{fig:aromacrowd}	
\end{minipage}
\hspace{5mm}
\begin{minipage}{0.5\linewidth}
	\centering
	\includegraphics[width=\linewidth]{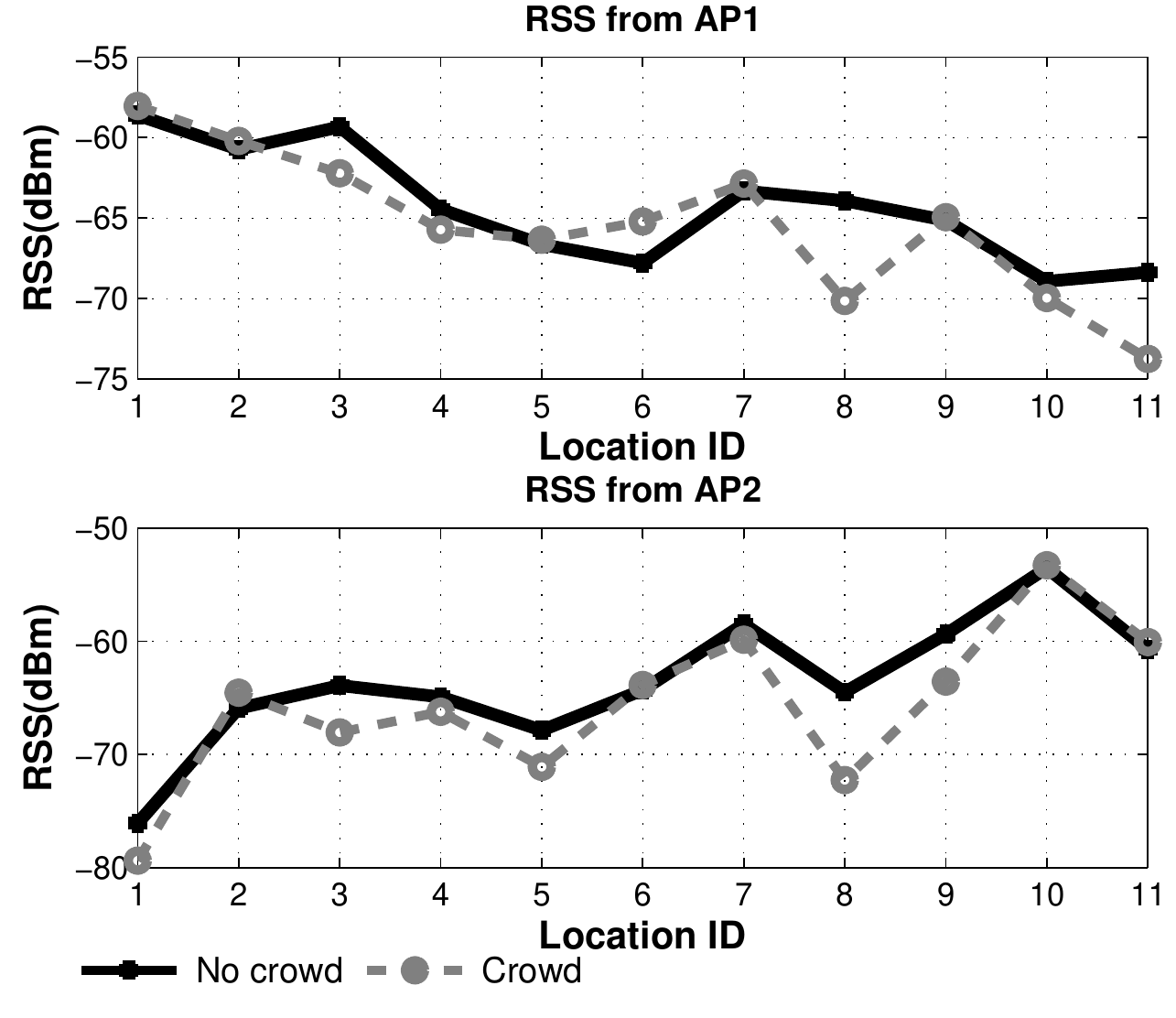}
	\caption{Effect of the {\bfseries party} crowd when using ceiling-mounted APs.}
	\label{fig:crowdceil}
\end{minipage}
\end{figure*}

Figure~\ref{fig:both2} shows the effect of the cluster scenarios (crowd
around APs) and Table~\ref{table:dbloc_acc} summarizes the localization accuracy results. The figure shows that a crowd around an AP mostly affect this AP only
and leads to a reduced RSS as compared to the no crowd
scenario. Since it is hard to construct the radio-map with
exactly the same crowd, it is usually the case that the radiomap
will be trained by a single person (no crowd). This leads
to a degradation of localization accuracy for the party crowd
as shown in Table~\ref{table:dbloc_acc}. The figure also shows that the most
affected locations, in terms of RSS, are those where a human
cuts the line-of-sight between the AP and this location.

We also study whether it is better to construct the radio-map with a single entity (no crowd) or a sample crowd scattered in the area of interest using ceiling or wall mounted APs (Figure~\ref{fig:aromacrowd}, Figure~\ref{fig:crowdceil} and Table \ref{table:dbloc_acc_1} ). A ceiling-mounted AP makes the signal interact with less bodies from the crowd and hence is closer to the no-crowd scenario,
as compared to a wall-mounted AP. This leads to a better localization accuracy in the case of ceiling-mounted AP as shown in Table~\ref{table:dbloc_acc_1}. The table also highlights that training with no crowd is better than training with a sample crowd, due to the change of the crowd between the training and operation phases.
\subsection*{Summary of Findings}
Through this section, our experiments lead to the following
conclusions:
\begin{itemize}
\item 	
Ceiling-mounted APs lead to higher localization accuracy.
\item
Using a higher frequency (e.g. 802.11a as compared
to 802.11b or g) leads to better localization accuracy.
However, due to reduced coverage, higher hardware installation
would be required.
\item 	
To accommodate for crowd scenarios, training with one
human is better than training with a crowd. In addition,
ceiling-mounted APs lead to better localization accuracy
in the case of crowd presence.
\end{itemize}

\section*{Device-Free Scenarios}\label{sec:df_scenarios}
In this section, we shift our attention to device-free localization, where entities are tracked without carrying any device based on their effect on the RSS at the infrastructure
devices. We study the following scenarios: effect of the devices mounting locations, upgrading the APs hardware (changing the operation frequency), and outside-entities effect. We end the section with a summary of our findings. We start by describing
the experimental testbed and evaluation metrics.
\begin{figure*}[!t]
\begin{minipage}{0.5\linewidth}
	\centering
	\includegraphics[width=\linewidth,height=6cm]{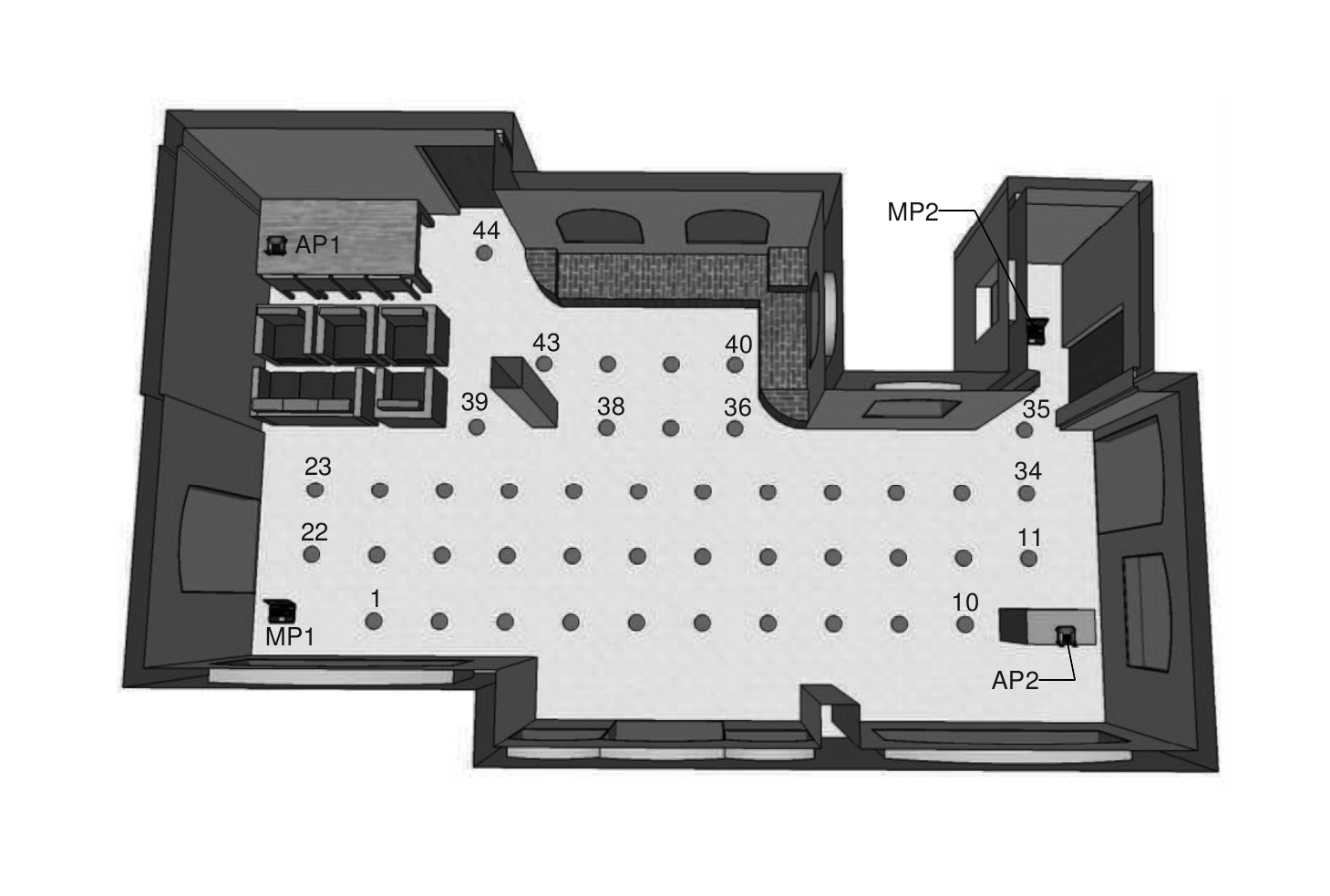}
	\caption{Device-free experiments layout. The figure highlights the locations of APs, MPs and radio map locations.}
	\label{fig:DeviceFree_Labeled}
\end{minipage}
\vspace{5mm}
	\begin{minipage}{0.5\linewidth}
	\centering
	\includegraphics[width=\linewidth]{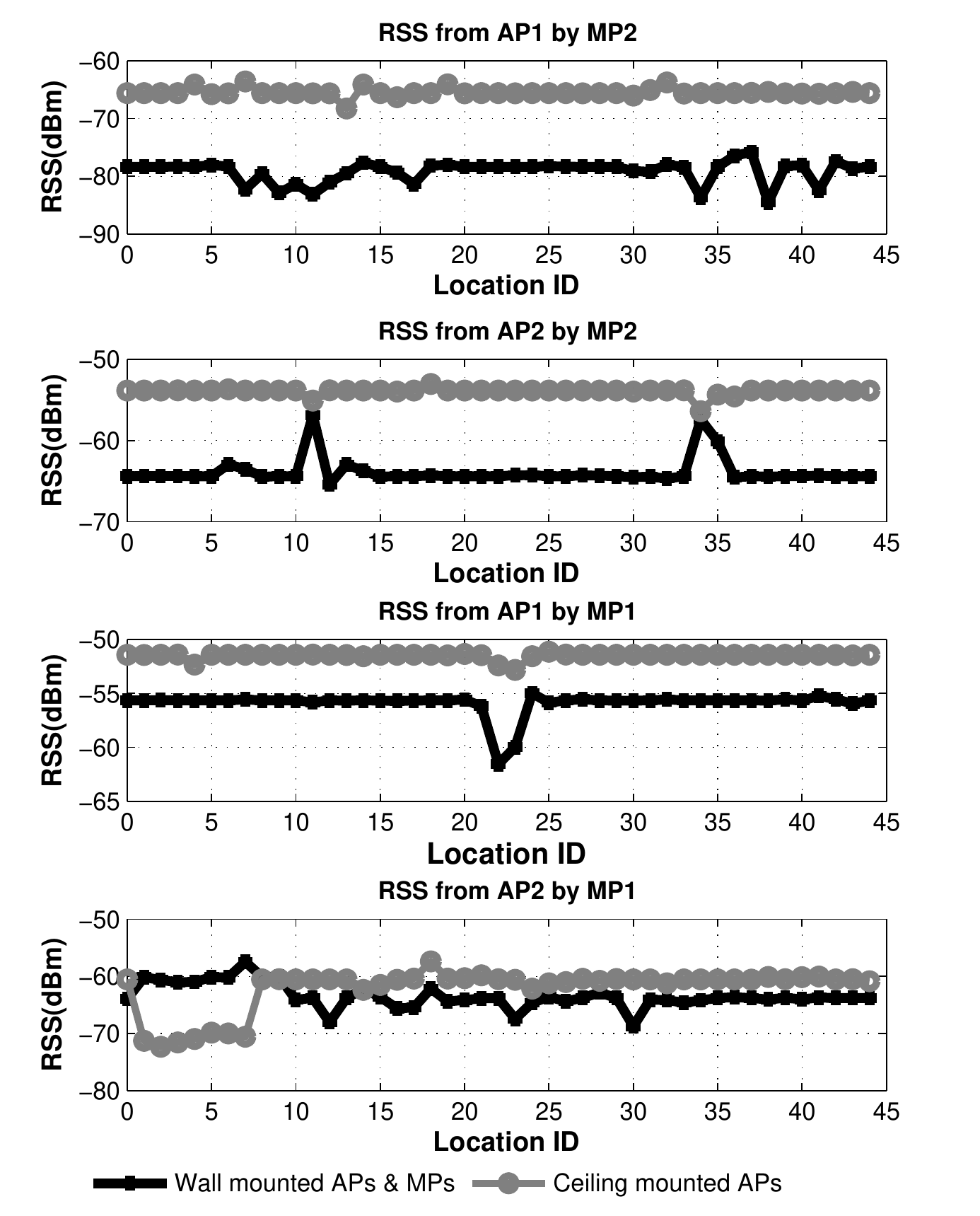}
	\caption{Effect of APs locations on the RSS from the four streams.}
	\label{fig:DF_height}
\end{minipage}
\end{figure*}
\subsection*{Experimental Testbed and metrics}
We used the same testbed and evaluation metrics as in
the device-based experiments. However, since the device-free
problem is more challenging, we used a denser radio-map of
forty four locations as shown in Figure~\ref{fig:DeviceFree_Labeled}. Note that the figure
also contains the locations of the laptops (i.e. the monitoring
points (MPs)) that are used as the infrastructure receivers. We
have a total of four RSS streams each corresponding to an
(AP, MP) pair. Location-zero in our results represents the
RSS values with no human in the area of interest (i.e. the silence
event).

\subsection*{Exp. 1: AP Mounting Location}

\begin{figure}[!t]	
	\centering
\includegraphics[width=0.7\linewidth,height=5cm]{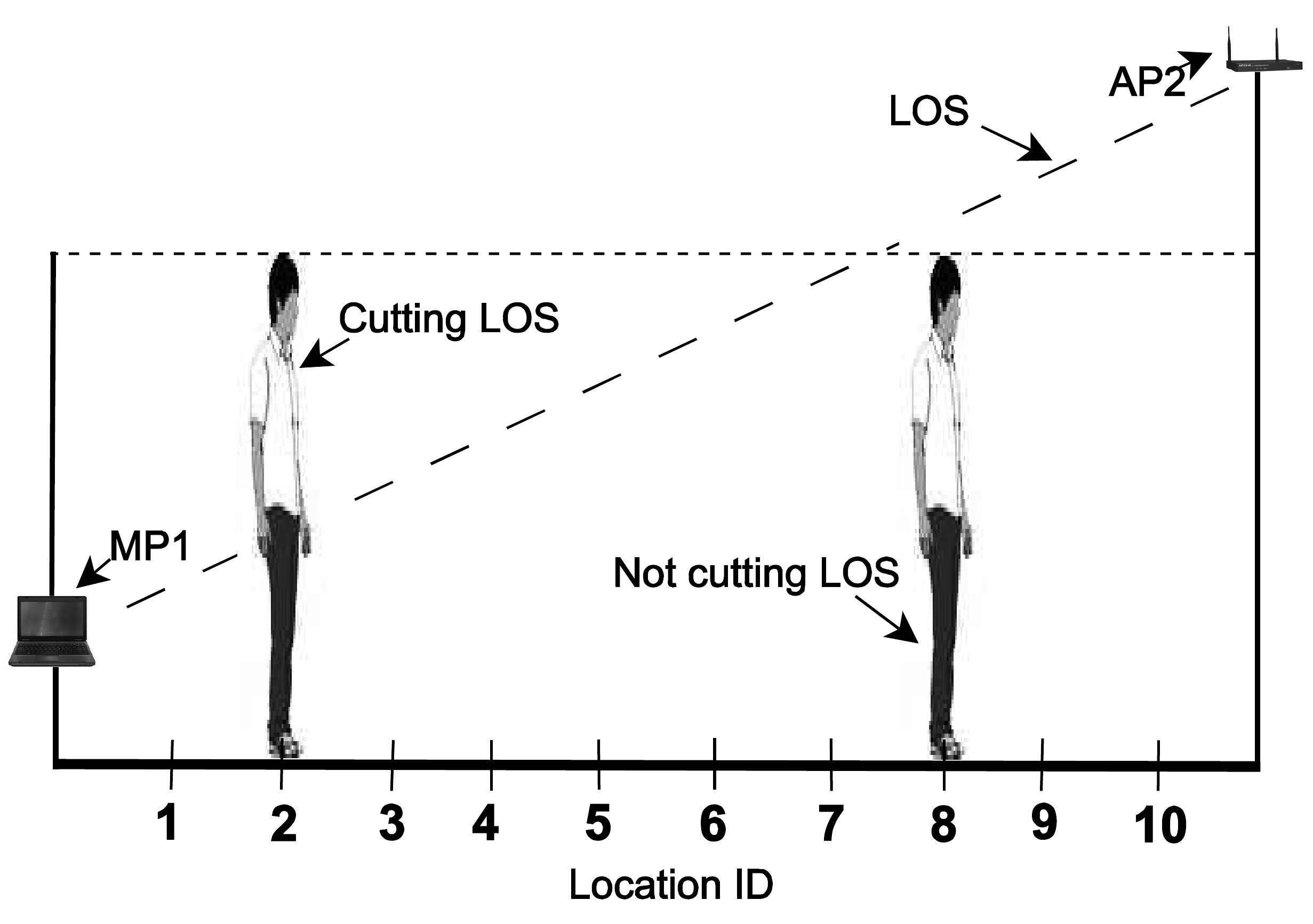}
	\caption{The human entity cuts the LOS between the AP and the MP only at
locations 1-7, explaining the attenuation pattern in the RSS stream from AP2
to MP1 (Figure~\ref{fig:DF_height}). AP2 was mounted to the ceiling and MP1 was at a height of $0.5m$. }
	\label{fig:AProof}
\end{figure}

\begin{table}[!t]
	\centering
		\caption{Average localization error in meters for the device-free scenario.}
	\begin{tabular}{|l|l|}
		\hline {\bfseries Experiment} & {\bfseries Acc.} (m)\\ \hline
		\hline {\bfseries Base exp.} (params. in Table~\ref{table:db_ap_config}) & 1.44\\
		\hline Exp. 1: Ceiling-mounted APs & 4.48\\
		 \hline Exp. 2: Freq. 5.7 GHz. & 1.77 \\
		\hline
	\end{tabular}
	\label{table:dfloc_acc_0}
\end{table}
For this experiment, we investigate the effect of using wall-mounted vs. ceiling-mounted APs. The laptops (MPs) where placed at a height of 0.5~m. Figure~\ref{fig:DF_height} shows the effect for the four streams. The figure shows that the locations that are most
affected are those that lie on the line-of-sight (LOS) between the AP and MP. For example, the most affected locations for
the RSS stream from AP2 to MP1 (Figure~\ref{fig:AProof}) are locations 1 to 7 because those are the locations where the entity cuts the LOS
between the AP and MP as shown in Figure~\ref{fig:AProof}. Therefore,
for the device-free case, wall-mounted APs lead to a higher chance of the entity cutting the LOS, as compared to ceiling mounted APs, and hence lead to better accuracy as quantified in Table~\ref{table:dfloc_acc_0}.

\subsection*{Exp. 2: Updating the AP H/W (Changing the Operating
Frequency)}
\begin{figure}[!t]
	\centering
	\includegraphics[width=0.6\linewidth]{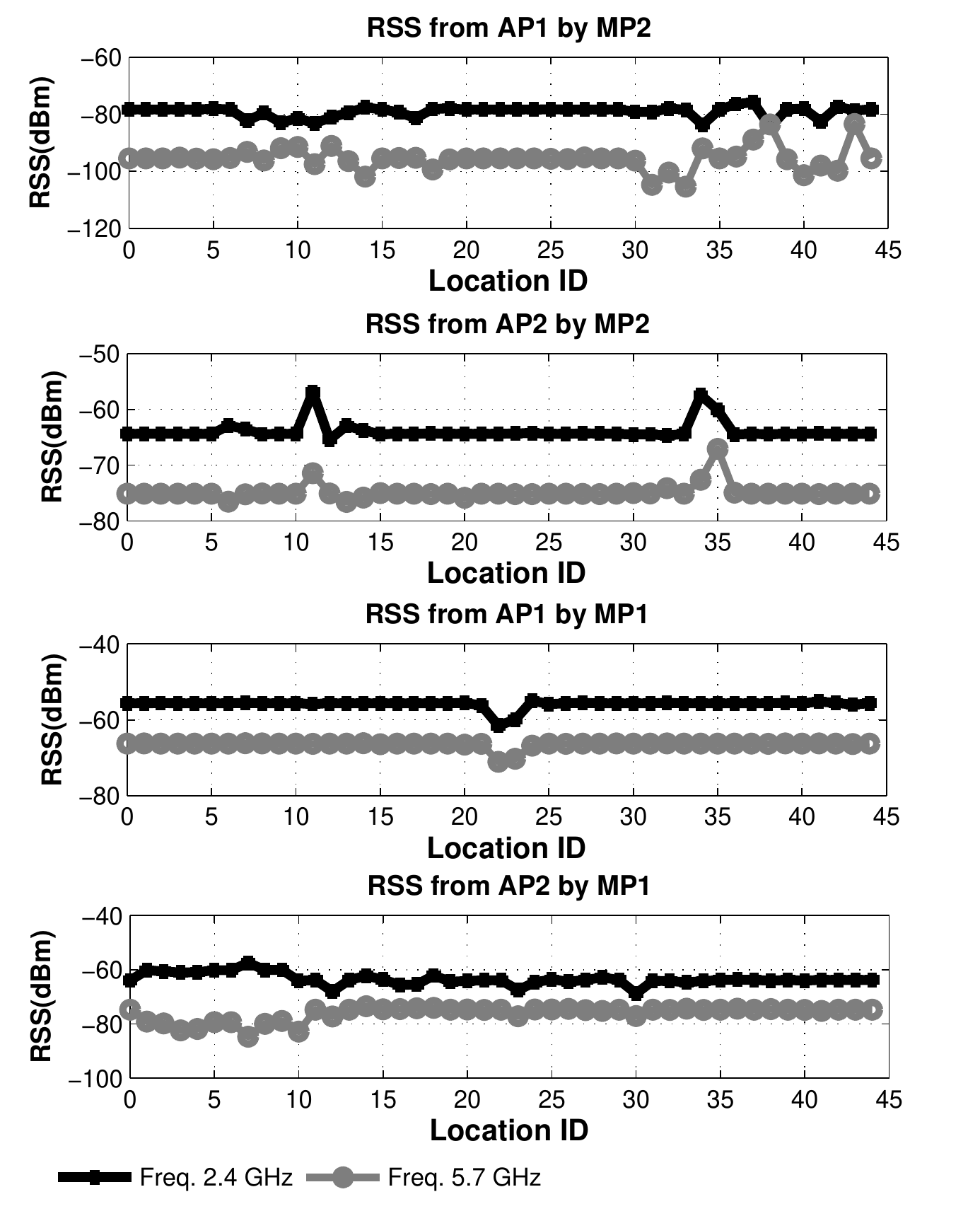}
	\caption{Effect of changing the frequency from the 2.4 GHz to 5.7 GHz on
RSS.}
	\label{fig:df_freq_aroma}
\end{figure}

In this subsection, we investigate the effect of switching the
frequency from the 2.4 GHz band to the 5 GHz band. Figure~\ref{fig:df_freq_aroma} shows the effect on the RSS while Table~\ref{table:dfloc_acc_0} shows the effect on the localization accuracy.\\
As expected, increasing the frequency to 5.7 GHz leads to more attenuation and hence a lower RSS. This leads to decreased RSS variance between the different locations and hence reduced localization accuracy.

\subsection*{Exp. 3: Outsiders Effect Scenario}
\begin{figure}[!t]
	\centering
	\includegraphics[width=0.7\linewidth]{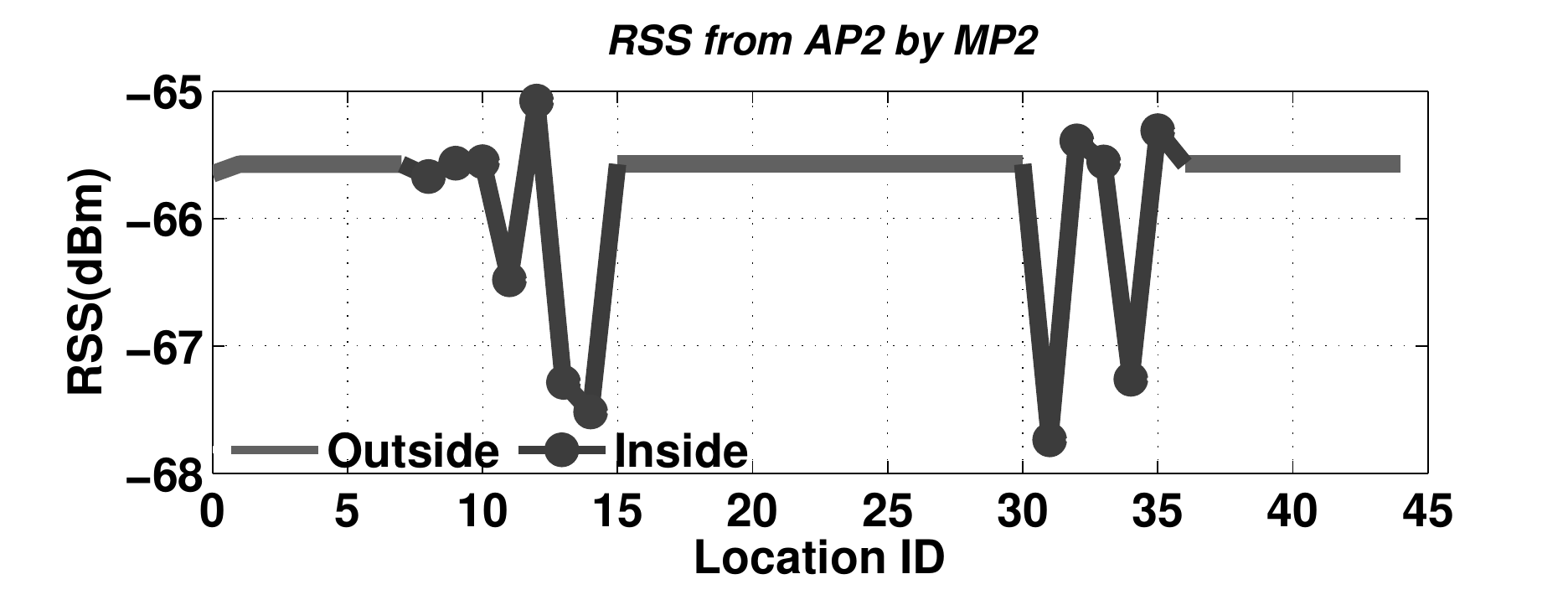}
	\caption{RSS values for the outside effect scenario.}
	\label{fig:inout}
\end{figure}

Through this scenario we aimed to study the effect of entities moving outside the area of interest on the RSS streams of WiFi hardware inside the area. This is a common scenario, e.g., a device-free system installed inside a platform and we want to study the effect of people moving outside it, in the street or neighboring apartments, on the system.
For this experiment, we closed the area on the far right of
the testbed (locations 8 to 14 and 31 to 35 in Figure~\ref{fig:DeviceFree_Labeled}) as the area of interest and we study the effect of a person moving at all
locations on the RSS of the stream (AP2, MP2), which is the only stream completely inside the area of interest. Figure~\ref{fig:inout} shows the results. The figure shows that when a person moves outside the area of interest, the RSS values are not affected.
This highlights the independence of different administrative entities and the promise of using device-free localization for large scale intrusion detection.

\subsection*{Summary of Findings}
Through this section, our experiments lead to the following
conclusions:
\begin{itemize}
\item 	
Contrary to the device-based case, wall-mounted APs lead to higher device-free localization accuracy.
\item 	
Similarly, using a higher frequency (e.g. 802.11a as compared to 802.11b or g) leads to lower localization accuracy.
\item 	
People moving outside the area of interest do not affect the streams that are completely inside the area of interest.
\end{itemize}

\section*{Conclusion}\label{sec:conclusion}
We explored different scenarios for device-based and device-free localization, highlighting factors that affect the localization process and showed how to tune them for better localization. Our recommendations from this study are: (1) for device-based localization: Use ceiling-mounted APs for better accuracy in tracking single devices and in party/crowd scenarios; the 5.7 GHz band (e.g. 802.11a) is preferred over the 2.4 GHz band in terms of accuracy. However, it needs higher number of APs to cover the same area; finally, calibration with
a single entity is better for tracking a crowd. (2) for device-free systems: Use wall-mounted APs for better accuracy; the 2.4 GHz band is preferred over the 5.7 GHz band for both the accuracy and coverage; Our analysis also shows that people moving outside the area of interest do not affect the device-free system inside the area. 

We believe that our analysis gives new insights for a wide range of entities interested in WiFi-based localization, both practitioners and researchers.

\bibliographystyle{elsarticle-num}
\bibliography{ms}

\begin{thebibliography}{10}
\expandafter\ifx\csname url\endcsname\relax
  \def\url#1{\texttt{#1}}\fi
\expandafter\ifx\csname urlprefix\endcsname\relax\def\urlprefix{URL }\fi
\expandafter\ifx\csname href\endcsname\relax
  \def\href#1#2{#2} \def\path#1{#1}\fi

\bibitem{radar}
P.~Bahl, V.~N. Padmanabhan, {RADAR}: an in-building rf-based user location and
  tracking system (2000) 775--784 vol.2.

\bibitem{horus}
M.~Youssef, A.~Agrawala, The horus {WLAN} location determination system, in:
  Proceedings of the 3rd international conference on Mobile systems,
  applications, and services, MobiSys '05, 2005, pp. 205--218.

\bibitem{Challenges}
M.~Youssef, M.~Mah, A.~Agrawala, Challenges: {D}evice-free {P}assive
  {L}ocalization for {W}ireless {E}nvironments, in: Proc. of the 13th annual
  ACM international conference on Mobile computing and networking, 2007.

\bibitem{nuzzer}
M.~Seifeldin, M.~Youssef, Nuzzer: A large-scale device-free passive
  localization system for wireless environments, CoRR abs/0908.0893.

\bibitem{rasid}
A.~E. Kosba, A.~Saeed, M.~Youssef, {RASID}: A robust {WLAN} device-free passive
  motion detection system, in: PerCom, 2012, pp. 180--189.

\bibitem{kalman}
M.~A. Seifeldin, A.~F. El-keyi, M.~A. Youssef, Kalman filter-based tracking of
  a device-free passive entity in wireless environments, in: Proceedings of the
  6th ACM international workshop on Wireless network testbeds, experimental
  evaluation and characterization, WiNTECH '11, 2011, pp. 43--50.

\bibitem{sscompwlan}
M.~Youssef, A.~Agrawala, Small-scale compensation for {WLAN} location
  determination systems, in: Wireless Communications and Networking, WCNC,
  Vol.~3, IEEE, 2003, pp. 1974--1978 vol.3.

\bibitem{kaemarungsi2004properties}
K.~Kaemarungsi, P.~Krishnamurthy, Properties of indoor received signal strength
  for {WLAN} location fingerprinting.

\bibitem{kaemarungsi2012analysis}
K.~Kaemarungsi, P.~Krishnamurthy, Analysis of {WLAN}'s received signal strength
  indication for indoor location fingerprinting, Pervasive and Mobile Computing
  8~(2) (2012) 292--316.

\bibitem{el2011impact}
K.~El-Kafrawy, M.~Youssef, A.~El-Keyi, Impact of the human motion on the
  variance of the received signal strength of wireless links, in: Personal
  Indoor and Mobile Radio Communications (PIMRC), 2011 IEEE 22nd International
  Symposium on, IEEE, 2011, pp. 1208--1212.

\bibitem{kosba2009analysis}
A.~E. Kosba, A.~Abdelkader, M.~Youssef, Analysis of a device-free passive
  tracking system in typical wireless environments, in: New Technologies,
  Mobility and Security (NTMS), 2009 3rd International Conference on, IEEE,
  2009, pp. 1--5.

\bibitem{aroma}
A.~Eleryan, M.~Elsabagh, M.~Youssef, {AROMA}: automatic generation of radio
  maps for localization systems, in: Proceedings of the 6th ACM international
  workshop on Wireless network testbeds, experimental evaluation and
  characterization, 2011, pp. 93--94.

\bibitem{aromaglobecom}
A.~Eleryan, M.~Elsabagh, M.~Youssef, Synthetic generation of radio maps for
  device-free passive localization, in: IEEE Globecom-Communication Software,
  Services, and Multimedia Applications Symposium, 2011.

\bibitem{wang2012no}
H.~Wang, S.~Sen, A.~Elgohary, M.~Farid, M.~Youssef, R.~R. Choudhury, No need to
  war-drive: Unsupervised indoor localization, in: Proceedings of the 10th
  international conference on Mobile systems, applications, and services, ACM,
  2012, pp. 197--210.

\bibitem{moussa2009smart}
M.~Moussa, M.~Youssef, Smart cevices for smart environments: Device-free
  passive detection in real environments, in: Pervasive Computing and
  Communications, 2009. PerCom 2009. IEEE International Conference on, IEEE,
  2009, pp. 1--6.

\bibitem{el2010propagation}
K.~El-Kafrawy, M.~Youssef, A.~El-Keyi, A.~Naguib, Propagation modeling for
  accurate indoor {WLAN} rss-based localization, in: Vehicular Technology
  Conference Fall (VTC 2010-Fall), 2010 IEEE 72nd, IEEE, 2010, pp. 1--5.

\bibitem{seifeldin2010deterministic}
M.~Seifeldin, M.~Youssef, A deterministic large-scale device-free passive
  localization system for wireless environments, in: Proceedings of the 3rd
  International Conference on PErvasive Technologies Related to Assistive
  Environments, ACM, 2010, p.~51.

\bibitem{kosba2012robust}
A.~E. Kosba, A.~Saeed, M.~Youssef, Robust {WLAN} device-free passive motion
  detection, in: Wireless Communications and Networking Conference (WCNC), 2012
  IEEE, IEEE, 2012, pp. 3284--3289.

\bibitem{sabek2012multi}
I.~Sabek, M.~Youssef, Multi-entity device-free {WLAN} localization, in: Global
  Communications Conference (GLOBECOM), 2012 IEEE, IEEE, 2012, pp. 2018--2023.

\bibitem{sabek2012spot}
I.~Sabek, M.~Youssef, Spot: An accurate and efficient multi-entity device-free
  {WLAN} localization system, Arxiv preprint arXiv:1207.4265.

\bibitem{kassem2012rf}
N.~Kassem, A.~E. Kosba, M.~Youssef, {RF}-based vehicle detection and speed
  estimation, in: Vehicular Technology Conference (VTC Spring), 2012 IEEE 75th,
  IEEE, 2012, pp. 1--5.

\bibitem{al2012rf}
A.~Al-Husseiny, M.~Youssef, {RF}-based traffic detection and identification,
  in: Vehicular Technology Conference (VTC Fall), 2012 IEEE, IEEE, 2012, pp.
  1--5.

\bibitem{abdel2013monophy}
H.~Abdel-Nasser, R.~Samir, I.~Sabek, M.~Youssef, Mono{PHY}: Mono-stream-based
  device-free {WLAN} localization via physical layer information, in: Wireless
  Communications and Networking Conference (WCNC), 2013 IEEE, IEEE, 2013, pp.
  4546--4551.

\bibitem{scholz2011challenges}
M.~Scholz, S.~Sigg, H.~R. Schmidtke, M.~Beigl, Challenges for device-free
  radio-based activity recognition, in: Workshop on Context Systems, Design,
  Evaluation and Optimisation, 2011.

\bibitem{hong2013ambient}
J.~Hong, T.~Ohtsuki, Ambient intelligence sensing using array sensor:
  Device-free radio based approach, in: Proceedings of the 2013 ACM conference
  on Pervasive and ubiquitous computing adjunct publication, ACM, 2013, pp.
  509--520.

\bibitem{sigg2013rf}
S.~Sigg, S.~Shi, Y.~Ji, Rf-based device-free recognition of simultaneously
  conducted activities, in: Proceedings of the 2013 ACM conference on Pervasive
  and ubiquitous computing adjunct publication, ACM, 2013, pp. 531--540.

\bibitem{chanproperties}
E.~C. Chan, G.~Baciu, S.~Mak, Properties of channel interference for {Wi-Fi}
  location fingerprinting.

\bibitem{UGTD}
D.~McNamara, C.~Pistorius, J.~Malherbe, Introduction to the uniform geometrical
  theory of diffraction (1990).

\bibitem{ghaddar2004modeling}
M.~Ghaddar, L.~Talbi, T.~Denidni, A.~Charbonneau, Modeling human body effects
  for indoor radio channel using {UTD}, in: Electrical and Computer
  Engineering, 2004. Canadian Conference on, Vol.~3, IEEE, 2004, pp.
  1357--1360.

\bibitem{seidel1994site}
S.~Seidel, T.~Rappaport, Site-specific propagation prediction for wireless
  in-building personal communication system design, Vehicular Technology, IEEE
  Transactions on 43~(4) (1994) 879--891.

\bibitem{aly2013new}
H.~Aly, M.~Youssef, New insights into wifi-based device-free localization, in:
  Proceedings of the 2013 ACM conference on Pervasive and ubiquitous computing
  adjunct publication, ACM, 2013, pp. 541--548.

\end{thebibliography}

\end{document}